# Magnetron plasma mediated immobilization of hyaluronic acid for the development of functional double-sided biodegradable vascular graft

Valeriya Kudryavtseva[1,a,b], Ksenia Stankevich[1,a,c], Anna Kozelskaya[a], Elina Kibler[a], Yuri Zhukov[d], Anna Malashicheva[d,e,f], Alexey Golovkin[e], Alexander Mishanin[e], Victor Filimonov[a], Evgeny Bolbasov[a,g*], Sergei Tverdokhlebov[a,*]

[a] National Research Tomsk Polytechnic University, Tomsk Russian Federation.
[b] Queen Mary University of London, London, United Kingdom.
[c] Montana State University, Bozeman, Montana, USA.
[d] Saint-Petersburg State University, Saint-Petersburg, Russian Federation.
[e] Almazov National Medical Research Centre, Saint-Petersburg, Russian Federation.
[f] ITMO University, Institute of translational Medicine, Saint-Petersburg, Russian Federation.
[g] V.E. Zuev Institute of Atmospheric Optics SB RAS, Tomsk, Russian Federation

[1] These authors have contributed equally to the study
* corresponding author

*E-mail address:*
ebolbasov@gmail.com
tverd@tpu.ru

**Abstract**

The clinical need for vascular grafts is associated with cardiovascular diseases frequently leading to fatal outcomes. Artificial vessels based on bioresorbable polymers can replace the damaged vascular tissue or create a bypass path for blood flow while stimulating regeneration of a blood vessel *in situ.* However, the problem of proper conditions for the cells to grow on the vascular graft from the adventitia while maintaining its mechanical integrity of the luminal surface remains a challenge. In this work, we propose a two-stage technology for processing electrospun vascular graft from polycaprolactone, which consists of plasma treatment and subsequent immobilization of hyaluronic acid on its surface producing thin double-sided graft with one hydrophilic and one hydrophobic side. Plasma modification activates the polymer surfaces and produces a thin layer for linker-free immobilisation of bioactive molecules, thereby producing materials with unique properties. Proposed modification does not affect the morphology or mechanical properties of the graft and improves cell adhesion. The proposed approach can potentially be used for various biodegradable polymers such as polylactic acid, polyglycolide and their copolymers and blends, with a hydrophilic inner surface and a hydrophobic outer surface.

**Graphical Abstract**

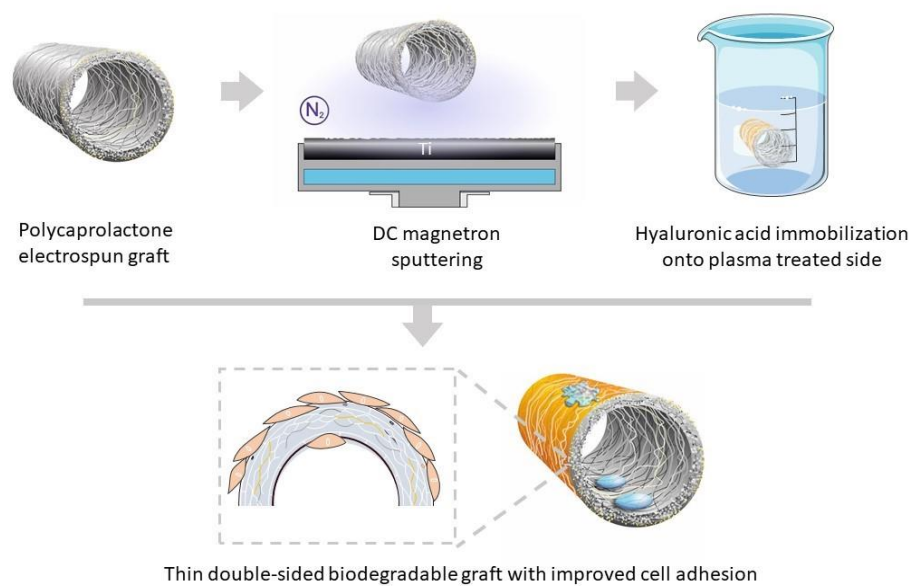

# 1. Introduction

Cardiovascular disease (CVD) accounts for the death of more than 18 million people a year, with vascular atherosclerosis being the cause in ~ 65 % cases [1]. The situation is aggravated by another acute problem arising from congenital malformations of large vessels (stenosis and atresia), when arteries or veins are narrowed, overgrown or completely absent due to intrauterine developmental defects. In most of these cases, the only possible treatment is to bypass the blood flow or directly replace the damaged (narrowed) vascular tissue with an artificial implant. Nowadays, due to significant clinical limitations, the use of auto-, allo-, and xenografts, as well as grafts made of biostable polymers (Gortex®, Dacron®), is possible only for prosthetics of large-sized vessels [1,2].

One of the promising strategies for solving the problem of graft deficiency is the development of artificial vessels based on bioresorbable polymers with an ability to stimulate regeneration of a blood vessel in situ [3]. Such vessels can be mass produced, ready to use, and could be easily commercialized. Due to their good physical and mechanical properties, high biocompatibility, low thrombogenesis and bioresorbability, electrospun artificial vessels made of polycaprolactone (PCL) are among the most promising. Among the other advantages of PCL are a lack of isomers, low melting point and good rheological and viscoelastic properties [4–6]. Despite the fact that elecrospun PCL scaffolds mimic the topology of extracellular matrix (ECM) and show good mechanical performance [7–9], they still suffer from high hydrophobicity which can affect cell adhesion [10,11]. Thus, the problem of proper conditions for the cells to grow on the inner volume of the vessel from the adventitia while maintaining its mechanical integrity of the luminal surface is one of the main challenges for vascular graft development [12].

To address this problem, several basic strategies have been proposed, including pore size gradient of graft in the direction of luminal surface-adventitia [13]; incorporation of biologically active compounds into the bulk or surface of the graft-forming fibers [14]; grafts based on composite fibers [15,16], plasma treatment [17] and etc. Despite the significant results achieved recently of each of the above strategies, none of them can be considered perfect. The strategy of gradient porosity is faced with the need to increase the thickness of the graft wall to prevent the risk of aneurysms, while significant difference in the wall thickness of the prosthesis could lead to changes in mechanical properties and cause graft failure [18]. Introduction of biologically active substances into the polymer spinning solution lead to changes in the electrical and rheological properties, leading to the formation of defective fibers and decrease in the graft material strength. Plasma treatment alters the wettability and increases the surface free energy, however this effect lasts several weeks which makes it difficult to use this strategy to create commercially available grafts with a long shelf life [19]. Therefore, the development of effective strategies for creating vascular grafts is considered to be an urgent task and the success of modern vascular surgery largely depends on its solution [20].

In this work, we propose a two-stage technology for processing vascular graft from PCL, which consists of 1) plasma treatment in a reactive magnetron discharge with the sputtering of a titanium target in a nitrogen atmosphere and 2) subsequent immobilization of hyaluronic acid on its surface. Plasma treatment enables the formation of a gradient coating in the bulk of the porous graft material, characterized by a negative concentration gradient of ions of the sputtered target from the external to the internal surface of the graft, while maintaining the graft structural integrity and mechanical strength [11,21]. The choice of reactive magnetron sputtering of a titanium target in a nitrogen atmosphere for processing PCL graft is due to formation of thin films of titanium oxide doped with nitrogen (TiON) on the surface of polymer material [22]. Such coating are able to produce nitric oxide [23], which accelerates the endothelization process [24]. However, an interesting phenomenon occurring in many plasma operating methods is hydrophobic recovery (or aging), which manifests itself as a gradual increase in water contact angle towards its original value [19].

Current strategies to reduce aging in plasma polymer film include enhancement of the degree of cross-linking, plasma coating architecture engineering, e.g. gradient structure and post-plasma grafting to decrease the number of reactive sites [25]. For most applications, especially biomedical, it is necessary to obtain a distinct number of functional groups. Therefore, plasma-based coatings with a vertical gradient structure have been proposed.

To obtain stabilized hydrophilic surfaces, a vascular graft modified in a plasma of a reactive magnetron discharge was kept in an aqueous solution of hyaluronic acid (HA) due to its high biocompatibility [26], high ability to maintain proliferation and normal functional activity of endothelial cells [27] and aortic smooth muscle cells (SMC) [28].

Hence, proposed strategy makes it possible to obtain a polycaprolactone vascular graft with a stable superhydrophilic outer surface (from the adventitious side) while maintaining high hydrophobicity of its inner surface (luminal surface).

## 2. Materials and Methods

### 2.1 Production of PCL vascular graft

To produce PCL vascular graft, 8% solution of PCL (80,000 g/mol, Sigma-Aldrich, UK) in trichloromethane (Fisher, UK) was prepared. The solution was stirred in a sealed glass reactor at 30°C until a homogeneous state then cooled to room temperature and loaded into an ES apparatus (NANON-01A, MECC Co., Japan) with the following technological parameters: polymer-solution flow rate - 6 mL/hr, needle-to-collector distance - 190 mm and voltage - 20 kV. The thickness of the obtained electrospun material was 236 ± 10 μm. For most of studies to obtain flat samples electrosun grafts were cut longitudinally.

### 2.2 DC Plasma Treatment

Prior to modification, the electrospun PCL vascular grafts were placed in a vacuum at $10^{-2}$ Pa to remove the residual solvent. The scaffolds were modified by a DC magnetron sputtering technique. The metal target was a chemically pure (99.99%) titanium (Ti) placed under a nitrogen atmosphere ($N_2$). The modification parameters were set as follows: the discharge power 20, 45, 75, 105, 135 W), operating pressure in the chamber of 0.7 Pa (99.99% $N_2$ gas), magnetron-to-target distance of 40 mm, magnetron area of 240 $cm^2$ and 4 min modification time. To avoid excessive rise of the sample temperature during the plasma treatment, the modification time was divided into one-minute treatments, each separated by a one-minute cooldown period. The maximum chamber temperature during the whole process was 39°C.

### 2.3 Immobilisation of HA

The plasma-treated samples were placed in a 0.1 wt.% aqueous solution of HA ($M_w = 2 \times 10^6$ g/mol, Sinopharm Chemical Reagent Co., China) for 30 min. The samples were washed with distilled water and dried at room temperature. The designed HA concentration (0.1 wt.%) maximises the biocompatibility of the scaffolds without evoking an additional immune response [29].

### 2.4 Scanning Electron Microscopy (SEM)

The morphology of the samples was investigated by SEM on an ESEM Quanta 400 FEG instrument (FEI, USA). Prior to the investigation, samples were coated with a thin gold layer by the magnetron sputtering system (SC7640, Quorum Technologies Ltd., UK). The fibre diameter was determined from SEM images captured in five fields of view using ImageJ 1.38 software (National Institutes of Health, USA). The average diameter was determined from at least 60 fibres.

### 2.5 Contact-Angle Measurements

The wettability of the samples was characterised by depositing 3 μL drops of polar liquid (water and glycerine) at different positions on the samples in a Krüss EasyDrop contact-angle measurement system and capturing the images one minute after depositing the drops. All data are represented as the averages and standard deviations of the measurements taken at five different spots on the surface of the respective sample.

### 2.6 X-Ray Photoelectron Spectroscopy (XPS)

XPS measurements were carried out in an Escalab 250Xi machine (Thermo Fisher Scientific Inc., UK) equipped with a monochromatic AlKα radiation source (photon energy: 1486.6 eV). The spectra were acquired in constant-pass energy mode at 100 eV for the survey spectrum and 50 eV for the element core-level spectrum. The spot size of the X-ray beam was 650 μm, and the total energy resolution was approximately 0.55 eV. Investigations were carried out at room temperature in an ultrahigh vacuum (with pressure of the order of $1 \times 10^{-9}$ mbar; in the electron–ion compensation system, the Ar partial pressure was $1 \times 10^{-7}$ mbar). The library of the reference XPS spectra, including the atomic registration sensitivity factors, was available in the Advantage Data System provided by the instrument manufacturer. The peaks were deconvoluted by Avantage software (Thermo Fisher Scientific Inc., UK) set to Shirley background subtraction followed by peak fitting to Voigt functions with an 80% Gaussian and 20% Lorentzian character. Each XPS experiment included 2 replicates. Each time, the measurements were done at least at 2 different locations.

## 2.7 Mechanical Studies

The mechanical properties of the samples were investigated by uniaxially stretching five pieces of each sample in a tensile testing machine (Instron 3369; Illinois Tool Works, USA) with a 50 N sensor. The traverse speed was set to 10 mm/min.

## 2.8 Cell adhesion studies

Given that the produced grafts had two different surfaces, the cell adhesion to each of them was studied. To do that, cells were separately cultured on the plasma-treated HA-coated side of the graft (further referenced as *PCL-HA top*) and the opposite, non-modified side of the graft (further referenced as *PCL-HA bottom*). Non-treated PCL graft samples were used as a control (Control samples).

Cell adhesion was studied using human multipotent mesenchymal stem cells (MMSCs) harvested from subcutaneous adipose tissue of healthy donors. All experiments were performed according to the Declaration of Helsinki within an approval of Ethics Committee of the Almazov National Medical Research Centre (no. 12.26/2014; December 1, 2014). Written informed consent was obtained from all the subjects before the fat tissue biopsy. Adipose-derived human multipotent MSCs had the following phenotype: CD19-, CD34-, CD45-, CD73+, CD90+, CD105+, as confirmed by flow cytometry (GuavaEasyCyte8; MerckMillipore, Darmstadt, Germany) and monoclonal antibodies (Becton Dickinson, Franklin Lakes, NJ, USA). Cells were cultured in α-

MEM medium (PanEco, Moscow, Russia) supplemented with 10% fetal calf serum (HyClone Laboratories, Inc., Logan, UT, USA), 1% L-glutamine, and 1% penicillin/streptomycin solution (Invitrogen, Waltham, MA, USA) in 37°C and 5% $CO_2$. Samples of the grafts were cut in 12 × 8 mm pieces and placed in the wells of a 24-well plate. Cells were seeded on top of the scaffold at a density of $5 \times 10^4$ cells per well, and co-cultured with the material for 72 hours in above mentioned conditions. The experiment was performed in triplicates. After 72 hours, the samples were transferred to the new wells of the plate, washed in PBS and fixed in a 4% solution of paraformaldehyde (PFA) for 10 minutes. The cells adhered to the sample surface were permeabilized by 0.05% Triton X-100 followed by rinse with PBS, blocked with 10% goat serum solution in PBS for 30 minutes at room temperature and incubated with an anti-vinculin antibodies (Thermo Fisher Scientific, Waltham, MA, USA) at 1:200 dilution for one hour. After three PBS washes, the cells were incubated with AlexaFluor 568 goat anti-mouse IgG (H+L) anti-bodies (Invitrogen, Waltham, MA, USA) at 1:1000 dilution for one hour at room temperature in the dark. The cells were washed thrice with PBS (5 min each) and stained with 4',6-diamidino-2-phenylindole (DAPI) for visualization of the nuclei.

For the quantitative and qualitative analysis, the adherent cells were imaged using Axiovert inverted fluorescence microscope (Zeiss, Germany) equipped with a Canon camera. Samples were placed between the two glass slides and ten different fields of view were captured at magnifications of ×10 and ×40 for each replicate. Quantitative analysis was performed by analysing images taken at a ×10 magnification (counting the nuclei of cells stained with DAPI), while qualitative analysis was performed using images taken at ×40 magnification (assessing the morphology of cells by stained cytoskeleton).

### 2.9 Statistical Analysis

Statistical analysis of physico-chemical data was performed in GraphPad Prism, version 8.00 for Windows (GraphPad Software, La Jolla California, USA) using Kruskal–Wallis test and Student's t test. The data are shown as mean (SD) ± standard deviation (SD). Statistical analysis of biological data was performed using the non-parametric Mann-Whitney U test. The data are presented as arithmetic mean (Mean) ± standard error (SE). Differences were considered significant at the $p < 0.05$ level.

# 3. Results and Discussion

SEM images of the PCL vascular grafts before and after plasma treatment are shown in Fig. 1.

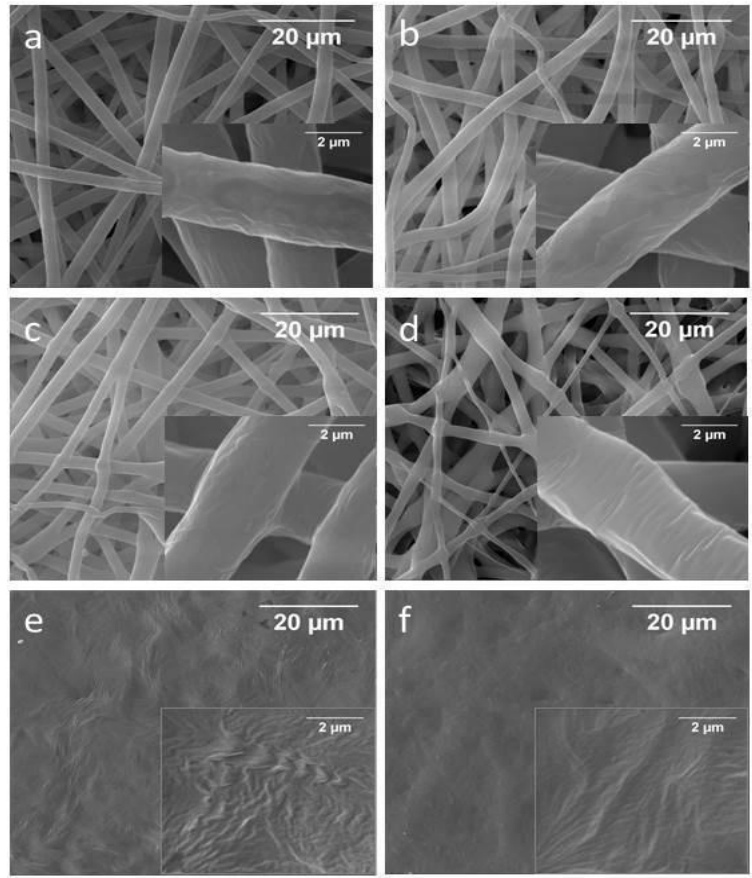

**Fig. 1.** SEM images of the electrospun PCL vascular grafts after plasma treatment at different currents: (a) untreated scaffolds and scaffolds treated at 20 W (b), 45 W(c), 75 W (d), 105 W (e) and 135 W (f).

The untreated electrospun PCL vascular graft consists of randomly intertwined cylindrical fibres with an average diameter of 2.0 ± 0.4 μm (Fig. 1(a)). As the treatment current increased from 20 to 45 A, the mean fibre diameter slightly decreased due to plasma etching and thermal impact. At a plasma treatment of 75 A, the fibres were melted in crossing areas and fibre surface was wrinkled. The formation of wrinkles and cracks on the fibre's surface could be caused by the difference in the elasticity of the polymer base and thin TiON coating whose thickness increases with an increase of the treatment time. An increase of discharge power leads to the overall melting of the fibres (Fig. 1 e-f).

The mechanical properties are essential for effective long-term implantation of tissue-engineered scaffolds. To evaluate plasma treated electrospun electrospun PCL vascular grafts, mechanical properties such as tensile strength, elongation and Young's modulus was measured. Results of the investigation of electrospun electrospun PCL vascular grafts mechanical properties are presented in Table 1.

**Table 1**. The mechanical properties of the electrospun PCL vascular grafts.

| Sample | Tensile strength, MPa | Young's modulus, MPa |
|---|---|---|
| Control | 3.54 ± 0.15 | 7.79 ± 0.32 |
| 20 W | 3.39 ± 0.32 | 10.44 ± 0.52* |
| 45 W | 4.02 ± 0.17 | 10.57 ± 0.35* |
| PCL-HA | 3.54 ± 0.30 | 10.83 ± 0.55* |
| 75 W | 7.01 ± 0.23* | 56.28 ± 17.07* |
| 105 W | 13.55 ± 0.19* | 159.14 ± 3.82* |
| 135W | 19.63 ± 2.63* | 171.42 ± 21.12* |

*$p < 0.05$ in comparison with control (Kruskal–Wallis test).

As the current increased in the plasma treatment, Young's modulus and tensile strength improved, especially in the 75-135W range. This trend can be attributed to thermal fibre bonding within the scaffold, which enhances the tensile properties of the electrospun PCL scaffolds by thermally induced shrinkage and molecular chain relaxation of the amorphous regions [7]. However, although the inter-fibre thermal bonding improved the mechanical stability of the electrospun scaffolds, it caused significant morphological changes (see Figs. 1(d)–1(f)).

Lowering the discharge power of the DC magnetron sputtering and applying the HA treatment slightly enhanced Young's modulus without changing the morphological properties (Figs. 1(a)–1(c)).

The effect of the plasma treatment and HA immobilisation on the wettability of the PCL nonwoven material was examined through contact-angle measurements. The results of the wettability investigation are presented in Fig. 2.

The contact angle was measured on water and glycerol droplets. The initial PCL scaffolds were hydrophobic with a water and glycerol contact angle of 122.4 ± 3.7° and 129.4 ± 2.6°, respectively (Figs. 2 and Fig. S3). Immediately after the DC plasma treatment at 20, 45 and 75 W, the top side of the scaffold was fully wetted while the bottom side remained hydrophobic (Figs. 2 and Fig. S3).

At the higher power (105 and 135 W), the bottom side became more hydrophilic: the contact angles of glycerol and water on the bottom sides were below 60° (Fig. S3). To demonstrate the differences in wettability of modified samples a highly viscous polar liquid, glycerol, was used.

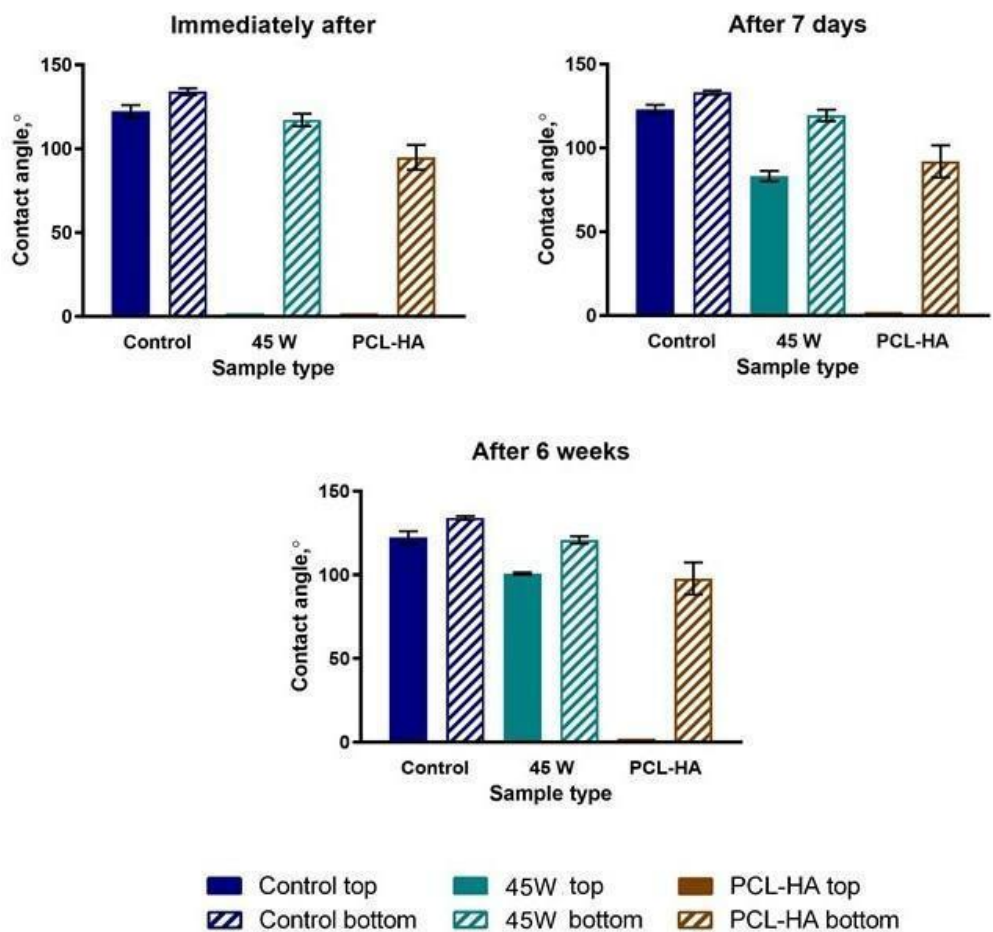

**Fig. 2.** Results of the wettability investigation.

To investigate the wettability changes over time, the contact-angle measurements of the DC plasma-treated samples were repeated after 3 days, 7 days and 6 weeks (Figs. S4-S6). A hydrophobic recovery of plasma treated surface of PCL scaffold was observed. For instance, the contact angle on the top side of the samples treated with 20 and 45 W increased to 38.4° and 41.8°, respectively, after 3 days and to 103.5° and 83.2°, respectively, after one week. However, no significant changes appeared on the bottom sides (Figs. S4-S5). The hydrophobic recovery was retarded on samples treated at a higher power (Fig. S5). After 6 weeks, all plasma treated samples exhibited hydrophobic properties on both sides, and the contact angles increased towards their original values (Fig. S6).

The HA immobilisation prevented hydrophobic recovery and the top side exhibited superhydrophilic properties throughout the study period (Figs. 2 and S7). As the treatment power increased, the bottom surface became more hydrophilic, as observed on plasma-treated scaffolds without additional modification. This effect is demonstrated in the video file in the Supplemental Data.

Considering the wettability, morphological features and mechanical properties of the produced materials, 20W was determined as the most suitable discharge power for modifying the electrospun PCL vascular grafts by the DC plasma treatment.

The effects of plasma treatment and HA immobilisation on the chemical composition and bonding states of the electrospun PCL vascular graft material were elucidated by XPS. Table 2 shows the atomic ratios and relative areas of the functional groups calculated by deconvoluting the C1s and N1s peaks.

**Table 2:** Relative areas corresponding to the chemical bonds and atomic compositions of the investigated electrospun PCL vascular grafts.

| Sample | Relative area corresponding to different chemical bonds, % | | | | | | Atomic ratio | |
| --- | --- | --- | --- | --- | --- | --- | --- | --- |
| | Carbon | | | | Nitrogen | | | |
| | C1 | C2 | C3 | C4 | $NH_2$ and/or $H_2NCO$ | $NH_3+$ | C/O | Ti/N |
| BE (eV) | 285.0 ± 0.2 | 285.6 ± 0.2 | 286.5 ± 0.2 | 289.1 ± 0.2 | 399.9 ± 0.2 | 402.5 ± 0.2 | | |
| PCL lit. [22] | 52 | 17 | 17 | 14 | - | - | - | - |
| Control | 54 ± 2 | 16 ± 1 | 16 ± 1 | 14 ± 1 | - | - | **3.17** | - |
| 20 W top (fresh) | 56 ± 2 | 7 ± 1 | 20 ± 1 | 17 ± 1 | 71.7 | 28.3 | **1.33** | **2.17** |
| 20 W top (6 weeks) | 55 ± 2 | 15 ± 1 | 14 ± 1 | 16 ± 1 | 68.0 | 32.0 | **1.08** | **2.91** |
| 20 W bottom | 54 ± 2 | 16 ± 1 | 16 ± 1 | 14 ± 1 | - | - | **3.15** | - |
| PCL-HA top | 53 ± 2 | 7 ± 1 | 23 ± 1 | 17 ± 1 | 77.2 | 22.8 | **1.21** | **2.92** |
| PCL-HA bottom | 54 ± 2 | 16 ± 1 | 16 ± 1 | 14 ± 1 | - | - | **3.16** | - |

The atomic constituents of the PCL vascular grafts give rise to unique spectral peaks in the photoelectron spectrum (e.g. the C1s peak), informing the bonding states of the PCL atoms. In the XPS survey spectrum of PCL vascular grafts peaks corresponding to C1s and O1s could be found.

The C1s spectra of the control PCL vascular grafts were consistent with those reported in previous studies [19] [30] [31] (see Table 2 and Fig. 4).

In the XPS survey spectrum of the fresh plasma treated PCL vascular grafts peaks corresponding to C1s, N1s, Ti2p and O1s could be found. The plasma treatment altered the shape of the C1s peak, increasing the C3 and C4 components and decreasing the C2 component. Peaks of amine and/or amide moieties ($NH_2$, $H_2NC=O$, 399.6 eV) and protonated/hydrogen bonded amine ($NH_3^+$, 402.3 eV) [31] groups appear in the N1s spectra of the plasma-treated PCL vascular grafts (Fig. 3). Enhanced surface chemical bonding states related nitrogen functional groups were also confirmed by the surface chemical composition, in which Ti and N appeared after the plasma treatment (Table 2).

In the Ti2p spectrum of the plasma-treated PCL vascular grafts three components corresponding to $TiO_2$ were observed: $Ti2_{p1/2}$ at 464.2 eV, $Ti2_{p3/2}$ at 458.3 eV, and satellite peak at 471.5 eV (Fig.4) [32]. The titanium chemical state is also confirmed by appearance of the $TiO_2$ peak at 529.8 eV in O1s spectrum (Fig. 5).

The deposition of $TiO_2$ instead of TiN after the plasma treatment could be explained by the following reasons. It is known that hydrophobic surfaces can adsorb water [33]. Considering the large surface of vascular grafts and low water desorption rate, even in vacuum [34] the titanium ions from the sputtering target could react with remaining water with the formation of $TiO_2$. Standard Gibbs free energy of formation for $TiO_2$ is -888.8 kJ/mol, whereas standard Gibbs free energy of formation for TiN is -243.8 kJ/mol [35]. Thus, the formation of $TiO_2$ is thermodynamically favoured over TiN.

The mechanism of the coating deposition on PCL-based grafts surface is probably similar to one described for polylactic acid [16]. The N2+ ions from plasma attack the PCL surface forming N· radicals and simultaneously breaking C-C, C-H and C-O bonds in PCL. The homolytic bond cleavage results in formation of various radical species on the polymer surface. These species can further react with nitrogen or residual water.

Thus, the interaction with nitrogen would result in amine and amide groups as was found by XPS. The incorporation of oxygen originating from water could give ether [–C–O–C–] as well as hydroxyl moiety [–(C–OH)–]. Indeed, the XPS study showed increased C3 and C4 components and decreased C2 component in the C1s spectra of the plasma treated PCL grafts. The oxidation of amino groups generated on the PCL surface to oximes and amides can also proceed resulting in increased oxygen content on the PCL surface.

After the plasma treatment, the polar side groups formed on the PCL grafts surface probably contributed to the hydrophilicity improvement of the scaffold, as observed in the contact-angle measurements (Fig. 2). The XPS spectra and elemental composition of the plasma treated PCL grafts bottom were like those of PCL control graft, indicating that modification did not occur on the other side (Table 2).

In the plasma-treated PCL vascular graft stored for six weeks, the component intensities of the C1s peak differed from those of the freshly treated PCL graft and resembled those of the control graft. The C3 component was decreased and the C2 component was increased comparing to freshly treated PCL graft. In N1s spectra the intensity of the component corresponding to amine and/or amide moieties decreased, whereas the protonated amine component intensity increased (Table 2). Also, the decreased C/O ratio observed (Table 2).

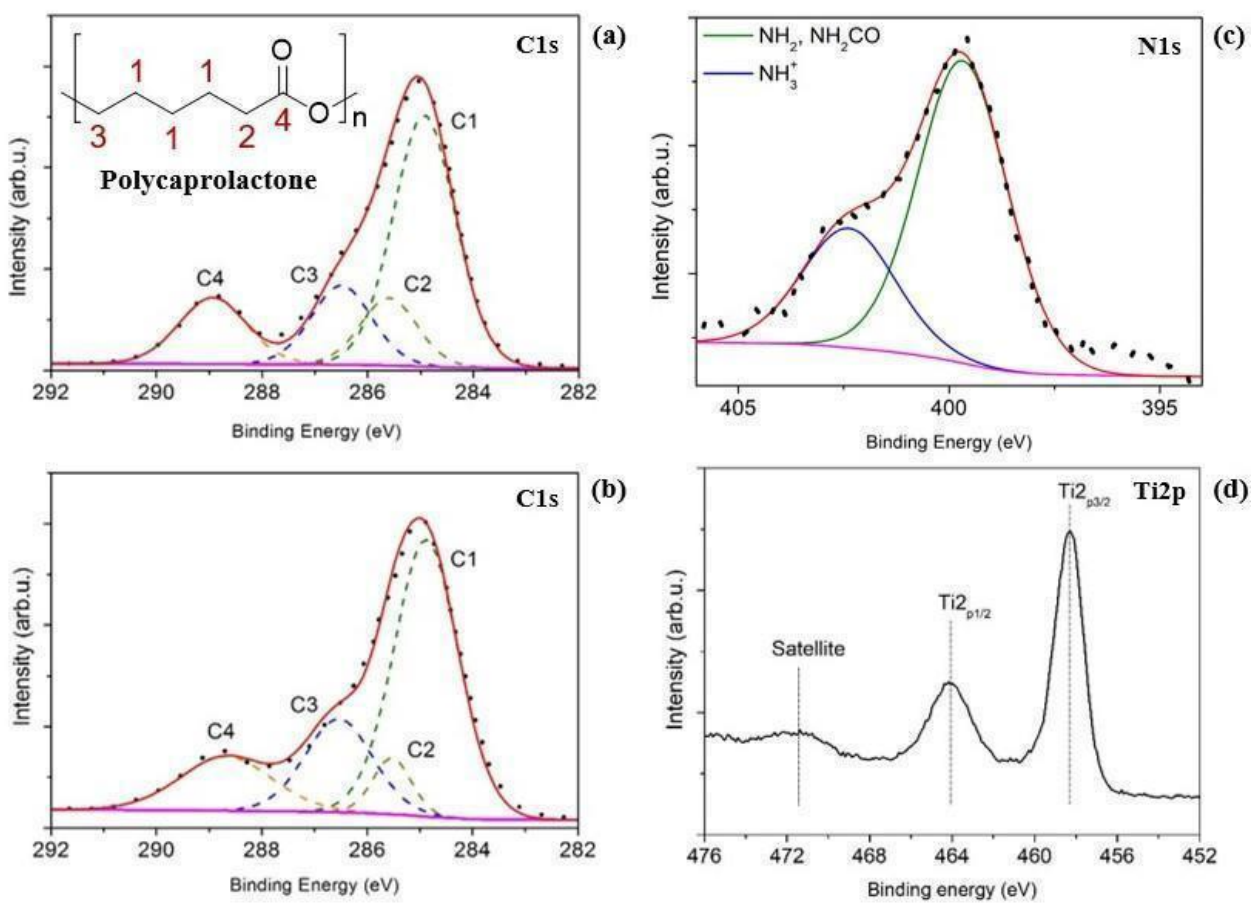

**Fig. 4.** C1s core-level spectra of the control PCL grafts (a); C1s core-level spectra (b), N1s core-level spectra (c) and Ti2p core-level spectra of the PCL grafts immediately after treatment with magnetron DC plasma at 20 W.

The changes in chemical bonding states and element content as well as hydrophobic recovery (Fig. 2) of the scaffold surface after six weeks could be explained by the mechanisms of polymer "aging" described earlier in [36]. The increase in oxygen and nitrogen amount could be caused by the adsorption of contaminants like water from the atmosphere on the scaffolds surface during the sample storage. The other possible processes explaining the changes in elemental composition and wettability of the scaffold are sublimation or inward-diffusion of low molecular weight oxidized material appearing during plasma treatment, reorientation of the polar side groups of the macromolecules towards the subsurface region of the material, and outward diffusion of low molecular weight oligomers or additives [36].

After HA immobilisation on the plasma treated PCL grafts surface, the intensity of the C3 components of the C1s line increased (Table 2). The intensities of peaks corresponding to amine/amide moieties and protonated amino group (Table 2) increase and decrease, respectively. This can be explained by the presence of [–(C–OH)–], [–C–O–C–], and [-C(O)-N] moieties in the HA chemical structure and is consisted with previous reports [37]. The observed changes in surface chemical bonding states (Table 2) confirm HA immobilisation on the PCL surface. Comparing to O1s spectra of plasma treated PCL large high-energy component corresponding to several groups including amide (531.6 eV), ether (532.6 eV), hydroxy (532.9 eV), carbonyl (532.3 eV) and acetal (533.1 eV) increases in the O1s spectra of the PCL-HA graft (Fig. 4) [37,38].
The largish proportion of polar groups in the HA structure contributed to the long-term superhydrophilicity of the PCL-HA surface (Fig. 2).

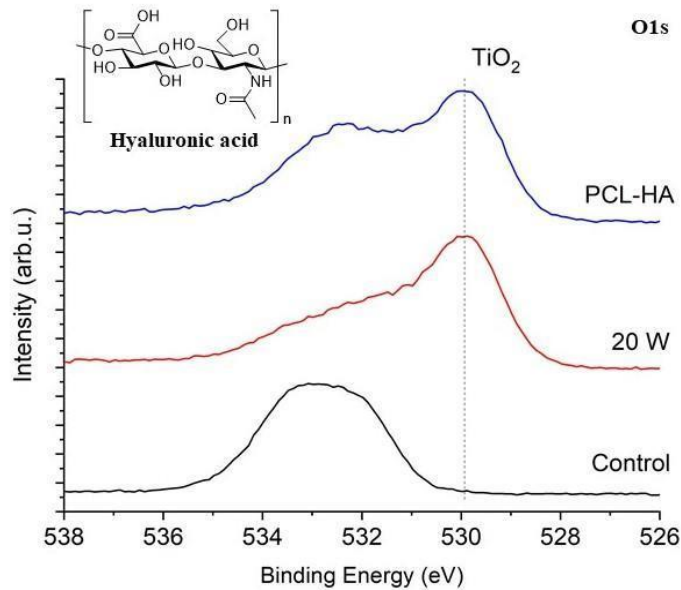

**Fig. 5.** O1s core-level spectra of the control PCL grafts, PCL grafts immediately after treatment with magnetron DC plasma at 20 W and PCL-HA grafts.

HA could bind to the PCL surface either covalently or by physical adsorption. It is known that if polymer is treated at vitreous state, the generated radicals and ions have reduced mobilities and can be trapped. Such trapped species could have a long lifetime. Thus, when a plasma treated graft was immersed in aqueous HA solution, the radicals remaining on its surface could interact with both water and HA resulting in covalent bonding of HA to PCL. However, the obtained data showed no strong evidences to support the described mechanism. On the other hand, partially protonated amino groups appear on the PCL surface after the plasma treatment. These groups can form a network of hydrogen bonding with hydroxy and carboxy groups of HA favouring HA physical adsorption on the graft surface. The elucidation of HA binding mechanism to plasma treated PCL graft surface is a complex issue that requires further additional studies.

The XPS spectra and elemental composition of the PCL-HA grafts bottom were like those of PCL control scaffold, indicating that HA was not immobilized on the other side (Table 2).

To assess the influence of plasma treatment and HA immobilisation on biological properties of the fabricated materials, we cultured human adipose-derived MSCs on the superhydrophilic and hydrophobic surfaces of the scaffold (Fig. 6) and evaluated the number of adhered cells and their morphology (Table 3).

Table 3. Number of adhered MSCs on the sample surfaces, cells/mm$^2$

| Sample | Mean±SE |
|---|---|
| Control | 118.7±4.7 |
| PCL-HA bottom | 160.6±6.9[*] |
| PCL-HA top | 200.6±7.1[*,**] |

[*] $p<0.0001$ comparing to control
[**] $p<0.05$ comparing to *PCL-HA bottom* and *PCL-HA top*

The fluorescence microscopy results have shown that all studied samples supported MSC attachment and growth *in vitro*. While some MSCs adhered to the surface of the material, the others started migrating into the bulk of the graft. On the control PCL sample as well as on the *PCL-HA bottom* sample isolated spindle-shaped cells were mostly found (Fig. 6). The number of MSCs on the surface of the PCL control scaffold was significantly lower compared to other samples ($p <0.0001$) (Table 3). It indicates a low functional activity of cells and a weak interaction of cells both among themselves and with the surface of untreated PCL graft [39]. The low functional activity of MSCs on the surface of PCL grafts without modification was previously observed in [40–42] due to the absence of chemically active functional groups on surface, which inhibit cell attachment, while the high hydrophobicity of polymer graft prevents the transport of dissolved nutrients and the removal of cell waste products.

Plasma treatment of the outer surface of the PCL vascular graft and subsequent immobilization of HA increased the number of adherent cells on the outer surface (*PCL-HA top*) by more than 65% compared to the surface of the untreated sample (Table 3). MSCs cultured on *PCL-HA top* samples were connected via numerous cytoplasmic bridges forming syncytium (Fig. 6g-i). Improved cell adhesion to the surface of *PCL-HA top* samples is most likely caused by hydrophilicity of the scaffold surface due to introducing the oxygen- and nitrogen containing polar groups via plasma treatment with subsequent immobilisation of HA. It is known that HA mediates various cellular events in vivo including adhesion and morphogenesis, mainly though the interaction with the cell surface receptor CD44 [43]. On the other hand, the common shortcoming of HA-based tissue engineering scaffolds is poor cell adhesion [43] that can be circumvented by combining HA with adhesion stimulating molecules such as gelatin [44]. In our case, the observed improved cell adhesion to *PCL-HA top* can be attributed to the complex hybrid character of the coating that supports both cell binding through both CD44-HA interaction and amino-group presence.

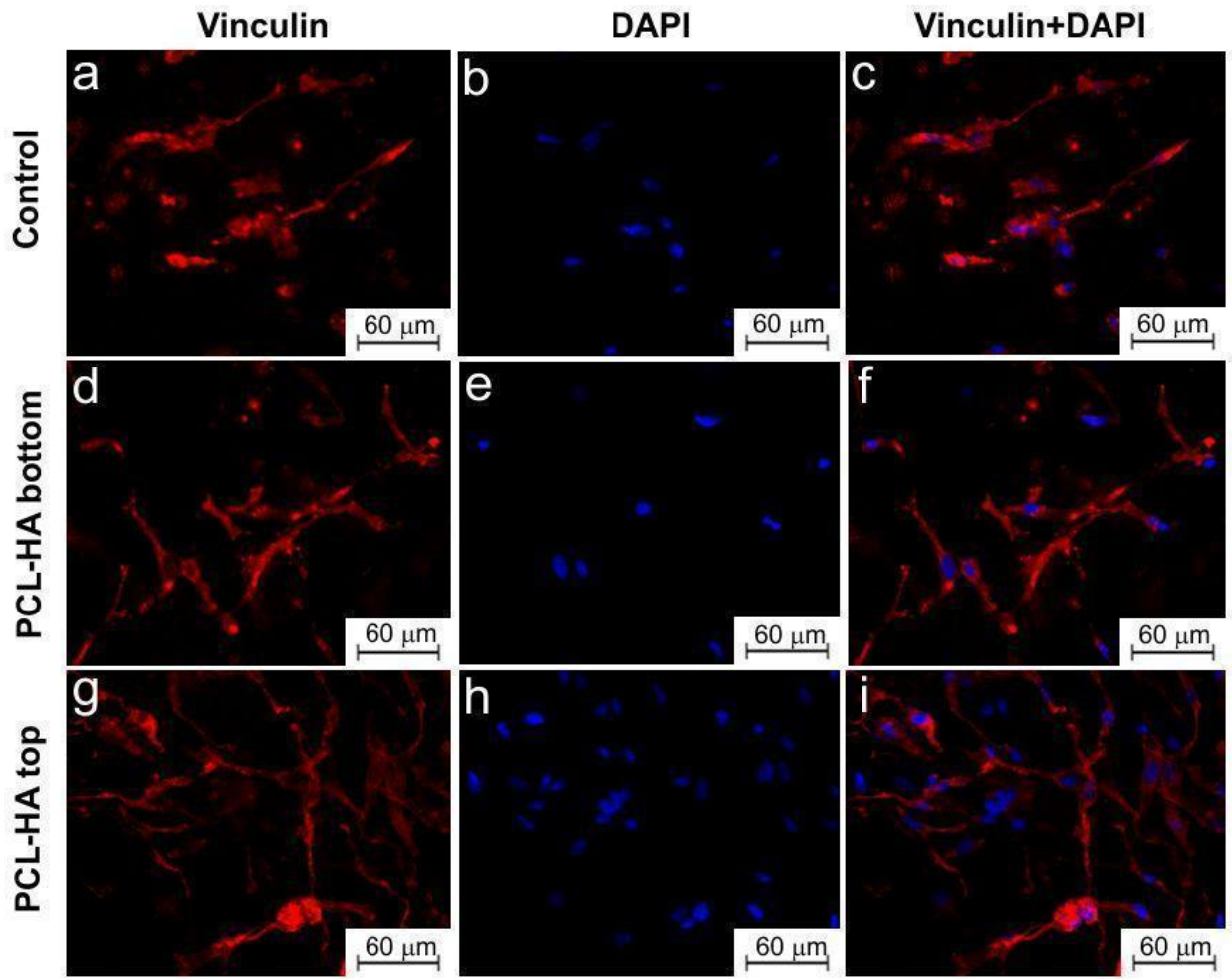

**Fig. 6.** Human adipose-derived MSCs stained on the surface of control PCL grafts (a-c), PCL grafts after immobilisation of HA (d-f) and PCL grafts after treatment with magnetron DC plasma at 20 W and immobilisation of HA (g-i). Vinculin is visualized in red (AlexaFluor 568 goat anti-mouse IgG (H+L) anti-bodies), cell nuclei are visualized in blue (DAPI). Representative images taken at ×40 magnification are shown.

The number of cells on the inner surface of the PCL of the vascular graft (*PCL-HA bottom*) is less than on the outer surface (*PCL-HA top*), but significantly higher than on the untreated graft (Table 3). Probably, the reason for this is a low number of functional groups on the inner surface of the vascular graft and hydrophobicity, which prevents the immobilization of hyaluronic acid as well as cell adhesion.

Improved cell adhesion, cells functional state and intercellular interaction observed *in vitro* on the outer surface of the modified graft (*PCL-HA top*), could result in increased endothelization rate from adventitia *in vivo*. Moreover, insignificant changes in the structure, mechanical properties and chemical composition of the luminal surface PCL of the graft will allow to maintain a low probability of blood clots inherent to grafts of this type [12,45,46].

## 4. Conclusions

Thin porous PCL-based double-sided grafts were fabricated by DC plasma treatment with subsequent immobilisation of HA. The plasma treatment changed the surface wettability of the PCL-based scaffolds from hydrophobic to superhydrophilic. However, after six weeks, the surface restored its hydrophobicity. The immobilised HA not only conferred high biocompatibility, but also stabilised the superhydrophilic surface against hydrophobic recovery.

The proposed modification does not significantly affect the mechanical properties of the electrospun PCL grafts. The fabricated scaffolds with a hydrophilic outer surface that ensures proper endothelization and hydrophobic inner surface that prevents early loss of mechanical integrity and lead to gradient of wettability and degradation rate. Plasma scaffold surface modification followed by immobilization of hyaluronic acid improves cell adhesion.

Overall, the use of DC magnetron plasma with appropriate modification parameters and subsequent HA immobilisation is a simple and scalable method for producing thin double-sided PCL vascular grafts with on hydrophilic and one hydrophobic side. This approach can potentially be used for modification of grafts from biodegradable polymers such as PCL, PLA, PGA and their copolymers and blends.

## Acknowledgements

This study was financially supported by the Ministry of Science and Higher Education of the Russian Federation (State Project "Science" № FSWW-2020-0011). The authors acknowledge the Resource Centre of Saint-Petersburg State University "Physical methods of surface investigation" for conducting XPS study.